\documentclass{article}
\usepackage{spconf,amsmath,graphicx,balance}
\usepackage{makecell}
\usepackage{multirow}
\usepackage{color}
\usepackage{hyperref}
\usepackage{url}
\usepackage{marginnote}

\newcommand{\mbf}[1]{\ensuremath{\mathbf{#1}}}

\newcommand{\mcl}[1]{\ensuremath{\mathcal{#1}}}

\newcommand{\msf}[1]{\ensuremath{\mathsf{#1}}}

\def\x{{\mathbf x}}

\title{AUDIO STYLE TRANSFER}

\name{Eric Grinstein, Ngoc Q.~K.~ Duong, Alexey Ozerov and Patrick P\'erez}
\address{Technicolor \\ 975 avenue des {C}hamps {B}lancs, CS 17616, 35576 Cesson S\'evign\'e, France \\
eric.grinstein@outlook.com,\{quang-khanh-ngoc.duong, firstname.lastname\}@technicolor.com}

\begin{document}
\maketitle

\begin{abstract}
``Style transfer'' among images has recently emerged as a very active research topic, fuelled by the power of convolution neural networks (CNNs), and has become fast a very popular technology in social media. This paper investigates the analogous problem in the audio domain: How to transfer the \emph{style} of a reference audio signal to a target audio \emph{content}? We propose a flexible framework for the task, which uses a sound texture model to extract statistics characterizing the reference audio style, followed by an optimization-based audio texture synthesis to modify the target content. In contrast to mainstream optimization-based visual transfer method, the proposed process is initialized by the target content instead of random noise and the optimized loss is only about texture, not structure. These differences proved key for audio style transfer in our experiments. In order to extract features of interest, we investigate different architectures, whether pre-trained on other tasks, as done in image style transfer, or engineered based on the human auditory system. Experimental results on different types of audio signal confirm the potential of the proposed approach.


\end{abstract}
\section{Introduction and related work}
\label{sec:intro}

Both visual texture synthesis, whose goal is to synthesize a natural looking texture image from a given sample, and visual texture transfer, which aims at weaving the reference texture within a target photo, have been long studied
in computer vision \cite{Efros:1999, Efros:2001}. Recently, within the era of convolutional neural networks (CNNs), these subjects have been revisited with great success. 
In their seminal work, Gatys \emph{et al.} \cite{DBLP:journals/corr/GatysEB15a} exploit the deep features provided by a pre-trained object recognition CNN to transfer by optimization the ``style" (mainly texture at various scales and color palette) of a painting to a photograph whose layout and structure are preserved. The result is a pleasing painterly depiction of the original scene. This is a form of example-based non-photorealistic rendering, which is now available in many social applications.  
Since then, the work of Gatys \textit{et al.} has sparked a great deal of research, see recent review on neural style transfer \cite{JingYFYS17}.


\begin{figure}[t]%
    \centering
    \includegraphics[width=0.67\linewidth]{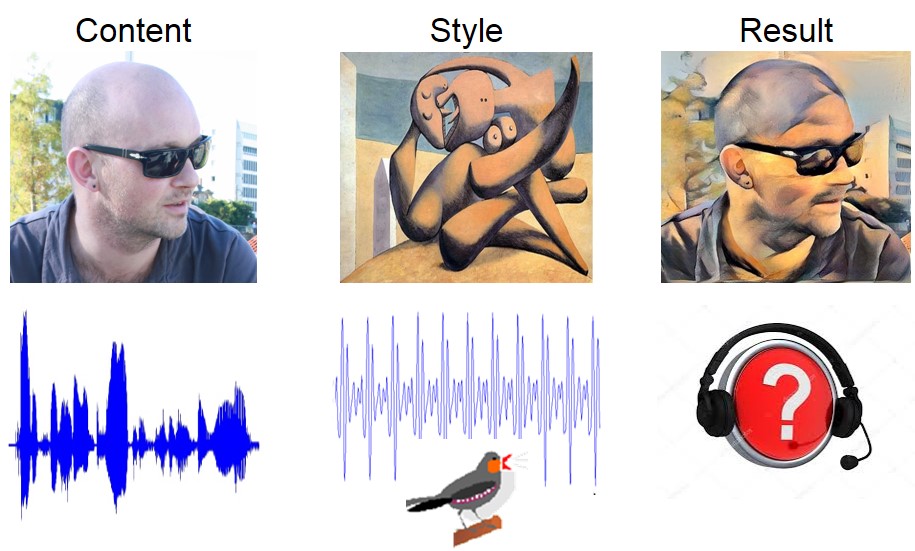} 
    \caption{{\bf Audio style transfer}: Drawing analogy from example-based image stylisation.}%
    \label{fig:P1}%
\end{figure}

Moving toward audio, \emph{sound texture} can be understood as the global temporal homogeneity of acoustic events \cite{MCDERMOTT2011926}, while the \emph{sound color palette} may be seen as a set of most representative spectral shapes. Note that model-based natural sound texture synthesis has also been extensively studied in the literature \cite{MCDERMOTT2011926, BrunaM13}. 
Closer to audio style transfer, a cross-synthesis method that hybridizes two sounds by simply multiplying the short term spectrum of one sound by a short term spectral envelop of another sound was proposed by Serra \cite{Serra}. More recently, Collins and Sturm proposed a dictionary-based approach for sound morphing, where sound textures can be represented by the dictionary's atoms  \cite{Collins2011}. To date, thanks to the power of CNNs, data-driven approaches have gained a lot in performance allowing high quality synthesis of complicated audio such as speech \cite{DBLP:journals/corr/OordDZSVGKSK16} and music \cite{DBLP:journals/corr/EngelRRDESN17}. Beyond texture synthesis and in analogy to image style transfer, some attention has been recently given to audio style transfer, notably in the form of exploratory work presented in blog posts \cite{ulyanov2016blog, foote2016blog}. These initial investigations showed that the original algorithm proposed in \cite{DBLP:journals/corr/GatysEB15a} for image is not fully adapted to audio signals. The result sounds often as a mixture of the style and content sounds \cite{ulyanov2016blog} rather than as a stylized content. Another attempt was done to apply style transfer for prosodic speech \cite{Perez2017}. The authors concluded that their approach, while allowing the transfer of low-level textural features, had a difficulty in transferring high-level prosody such as emotion or accent. 
Finally, regarding music, it is worth to mention a pioneering work by Pachet \cite{pachet2016joyful}, where the same music piece is automatically re-orchestrated in different styles. However, this re-orchestration is performed in the domain of symbolic music description (music scores) and not in the raw audio domain, as we consider here.


While visual style transfer provides new means to create compelling pictures for professionals and consumers, a success in audio signals would similarly open ambitious applications including professional audio editing, music creation, sound design and movie post-production (including dubbing). In this paper we present our contribution to this important new topic. In particular, we adapt the framework of Gatys \textit{et al.} with a new way of initializing the optimization process and a simplified loss (style only).  To extract the style statistics and to define the loss accordingly, we investigate the use of different CNN architectures, inspired by those used in image style transfer but adapted to audio signals. In addition, we explore the use of an auditory-motivated sound texture model. 

The rest of the paper is organized as follows. Section \ref{sec:discussion} discusses the audio style transfer problem together with our view on what could be audio \emph{style}. Section \ref{sec:framework} presents the proposed framework, which allows the use of different audio texture models and associated optimization-based synthesis algorithms in a flexible way. Experimental results are discussed in Section \ref{sec:experiment}. Finally, we conclude in Section \ref{sec: conclusion}.

\section{Problem setting and discussion}
\label{sec:discussion}


Gatys \emph{et al.} \cite{DBLP:journals/corr/GatysEB15a} introduced the concept of neural style transfer between images, with impressive results on transfering colors and paint strokes from a painting to a photograph. In this framework, starting from a \emph{content} image and a \emph{style} image, the corresponding deep feature maps related to content and to style are extracted and stored. The problem consists in computing a new image that matches well the content features and relevant statistics of style features. Starting from a random white noise image, this is obtained through iterative minimization of a suitable two-fold loss.     
As a result, the final image retains the global structure of the content image while borrowing textures and colors from the style image. 

A preliminary work on audio style transfer by Ulyanov and Lebedev \cite{ulyanov2016blog} relies on a similar optimization framework,
though using a wide, shallow, random network (a single layer with 4096 random filters) rather than a deep pre-trained one as in visual transfer.    
More specifically, a 1D audio waveform is transformed into a 2D representation using the short time Fourier transform (STFT). The resulting spectrogram representation, when discarding phase information, can be viewed as a 2D image for processing. However, it is still processed as a 1D signal of vector (frequency) observations. In other words, the color channels in image case are replaced by frequency channels. The spectrogram to be computed is initialized as random noise and iteratively updated to minimize a loss function between its features and those extracted from the content and style sounds. The results obtained with this method are still limited as the preservation of content and style remains unclear.

Note that though the notions of style and content are not formally defined and are context-dependent (depending on both the task and the data), researchers working on visual style transfer seem to agree on the following: ``Style'' refers mostly to the space-invariant intra-patch statistics, \textit{i.e.}, to the texture at several spatial scales, and to the distribution of colors a.k.a. the color palette; ``Content'' encompasses the broad structure of the scene, that is, its semantic and geometric layout 
\cite{DBLP:journals/corr/GatysEB15a, JingYFYS17}. In audio, the notions of style and content are even harder to define and would depend more on the context. For speech for instance, content may refer to the linguistic information like phonemes and words while style may relate to the particularities of the speaker such as speaker's identity, intonation, accent, and/or emotion. For music, on the other hand, content could be some global musical structure (including, \textit{e.g.}, the score played and rhythm) while style may refer to the timbres of musical instruments and musical genre. Fig. \ref{fig:P1} depicts the analogy between image style transfer explored in computer vision (first row),
and audio style transfer yet to be explored (second row).

\section{Proposed framework}
\label{sec:framework}

\begin{figure}%
    \centering
    \includegraphics[width=0.85\linewidth]{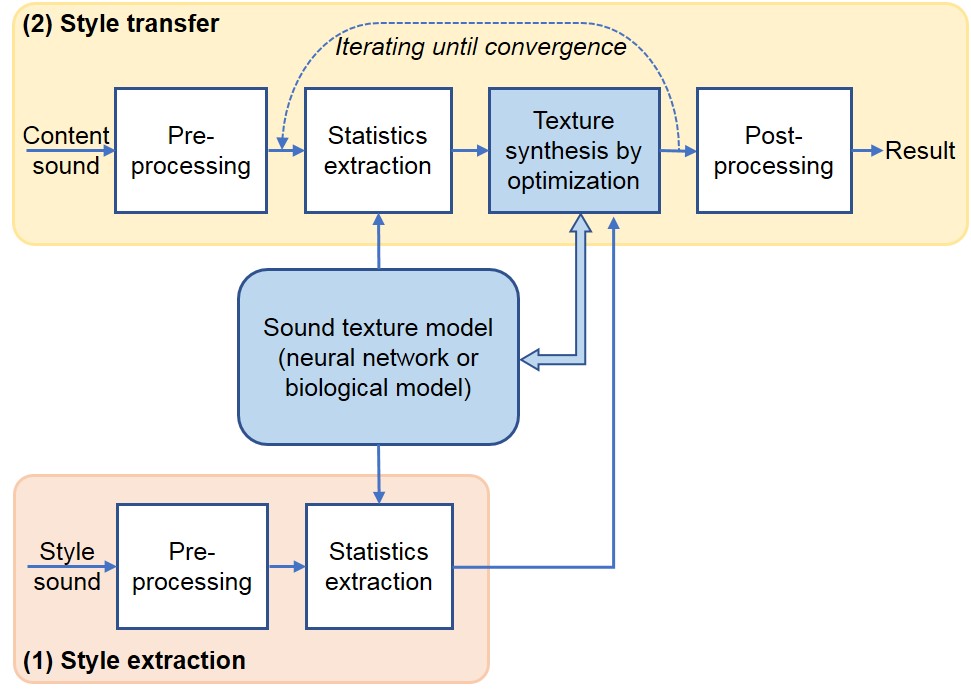} 
    \caption{\textbf{Proposed audio style transfer framework}: Given an audio texture extraction model (artificial neural net or auditory model), the \textit{content} sound is iteratively modified such that its audio texture matches well the one of the \textit{style} sound. If required by texture model, raw signals are mapped to and from a suitable representation space by pre/post-processing.}%
    \label{fig:NN}%
\end{figure}

A general workflow of the proposed audio style transfer framework is depicted in Fig. \ref{fig:NN}. It consists in two stages: (1) Style extraction and (2) Style transfer. In more detail, first the content and style waveforms are pre-processed, if required by subsequent steps, so as to obtain a desired signal representation. As an example, they can be transformed into 2D spectrogram representations by the STFT. Then the texture statistics of these signal representations are extracted by a sound texture model, which can be, \emph{e.g.}, a neural network or an engineered perceptual model. Given these texture statistics, an optimization algorithm is used to synthesize the representation of the final signal. If this representation is not in the waveform domain, a final post-processing is required to recover and play back the newly synthesized audio signal. 


Note that this framework differs from existing optimization-based visual and audio style transfer approaches \cite{DBLP:journals/corr/GatysEB15a, deepphoto17, ulyanov2016blog} in two related ways: It does not make use of a content loss and it does not initialize the signal to be optimized with a random noise. Instead, we propose to manipulate directly the content sound representation so as it gradually incorporates style texture into it. In other words, the new audio is initialized by the content sound and is modified to minimize a style loss only. We found this to be a key factor for producing compelling results, after numerous trials with random noise initialization and two-fold losses, following existing approaches. We note that despite the absence of content loss, the global structure of the content is nonetheless preserved in the final signal. Started from the content, the iterative optimization converges to a local minimum with similar structure. This preservation of the structure could be further enforced with a content loss, but it did not appear necessary in our experiments while being potentially harmful (making gradient descent less able to move away from content initialization). 
 
In the following, we present two types of sound texture models that we investigated.

\subsection{Neural network-based approach}
\label{ssec:NN}

Motivated by the works in image style transfer, we investigated the use of neural networks for audio style transfer.


Given the representation $\x$ of an input audio signal (raw signal or spectrogram), a CNN is used as follows to extract statistics that characterize stationary sound textures. Denoting $\msf F_\ell = [\mbf f_{\ell,k}]_{k=1}^{K_{\ell}}$ the matrix of the $K_{\ell}$ (vectorized) activation maps at layer $\ell$ of the network, Gatys \textit{et al.} proposed to use as statistics the Gram matrices
$\msf G_\ell = \msf F_\ell^\top \msf F_\ell$ at several layers $\ell\in L$ of the network.\footnote{Equivalently, this Gram matrix is, up to a factor and a shift, the space-invariant cross-channel covariance matrix of activations.} The style loss to minimize is: 
\begin{equation}
\mcl L(\mbf x; \mbf x_{\text{sty}}) = \sum_{\ell\in L}\big\|\msf G_\ell(\mbf x) -  \msf G_\ell(\mbf x_{\text{sty}})\big\|_F^2,
\label{eq:loss}
\end{equation}
where $\mbf x_{\text{sty}}$ is the style audio. We minimize this loss by gradient descent started at the content audio $\mbf x_{\text{cont}}$. Each gradient step requires to back-propagate from concerned activations all the way to network input $\mbf x$. With this approach, we tested several neural nets.  



\smallskip
\noindent{\bf VGG-19 \cite{Simonyan14c}}\quad This is a well-known deep neural network designed and trained for image classification. Gatys \textit{et al.} used it to extract features for image style transfer. As a 2D spectrogram can be viewed as an image, we first tried the VGG-19 architecture directly, taking features at the output of the five convolutional blocks, to see how it performs for audio style transfer. In this case $\x$ is obtained by transforming the 1D waveform into a 2D spectrogram representation by the STFT. In order to adapt to the input format of VGG, we replicate the spectrogram three times to obtain an RGB-like image. After optimization, we averaged the three RGB channels to obtain the final spectrogram. The missing phase information is estimated by Griffin \& Lim's algorithm \cite{griffinlim}, yielding the final sound waveform. 

\smallskip    
\noindent{\bf SoundNet \cite{DBLP:journals/corr/AytarVT16}}\quad This is a fully convolutional deep network that is specifically designed and learned on a large amount of unlabeled videos (including sounds). The resulting sound features have shown state-of-the-art performance on standard benchmarks for acoustic scene/object classification. With this architectures, $\x$ is a raw waveform and Gram matrices are computed for all 8 convolutional layers.
    
\smallskip
\noindent{\bf Wide-Shallow-Random network}\quad Recent work on image style transfer showed that untrained shallow neural nets can be used to extract style statistics \cite{DBLP:journals/corr/UstyuzhaninovBG16,DBLP:journals/corr/PuyKP17,DBLP:journals/corr/HeWH16}. This idea was also considered for audio in the preliminary work by Ulyanov and Lebedev \cite{ulyanov2016blog}. We use the same setting, one-layer CNN with 4096 random filters and convolution only performed along the temporal axis, taking 2D spectrograms as input.

\subsection{Auditory-based approach}
\label{ssec:mcdermott}

In the CNN-based approach, the sound texture model is data-driven, at least in case of VGG and SoundNet, which are pre-trained. Though CNNs are very powerful in general, it is quite difficult to interpret the resulting statistics. Thus we investigate an alternative approach where the sound texture model has been designed by 
experts of auditory perception. 
Our approach strongly relies on the sound texture analysis and synthesis system proposed by McDermott and Simoncelli \cite{MCDERMOTT2011926}. This texture model emulates the human auditory system through three sound processing steps (which can also be viewed as three layers of a hand-crafted neural network): cochlear filtering, envelope extraction and compressive nonlinearity, and modulation filtering. The first layer uses 30 bandpass cochlear filters to decompose the waveform into acoustic frequency bands. The second layer extracts the envelope of each frequency band computed in the first step and applies a compressive nonlinearity to it. 
The third layer further decomposes each compressed envelope by 20 bandpass modulation filters. These modulation filters are argued to be conceptually similar to coachlear filters, except that they operate on the compressed envelopes rather than on the sound waveform as in the first decomposition layer. 

Within this texture model, statistics used for capturing audio style are as follows (see \cite{MCDERMOTT2011926} for details): Variance of each coahlear filter band in the first layer; Mean, variance, skewness of each envelope band and cross-band correlation in the second layer; Power from each modulation band and cross-band correlation in the third layer. These statistics were shown to be suited for summarizing stationary sounds like textures over a moderate time scale. For synthesizing the sound signal from desired statistics, a variant of gradient descent is performed, starting from random noise signal. It minimizes a loss analogous to (\ref{eq:loss}) but with additional moments to match beside the variances and correlations captured by Gram matrices. 
\cite{MCDERMOTT2011926}. Our framework shown in Fig. \ref{fig:NN} proceeds similarly, using statistics of style audio as goal, but starting from the content sound so that its statistics progressively become closer to the desired ones.
Note that in this approach, the waveform signals are processed directly, 
thus the pre-processing and post-processing blocks in Fig. \ref{fig:NN} are not needed as in the case of SoundNet.

\section{EXPERIMENTS}
\label{sec:experiment}

\begin{figure}[t]
    \centering
    \includegraphics[width=0.9\linewidth]{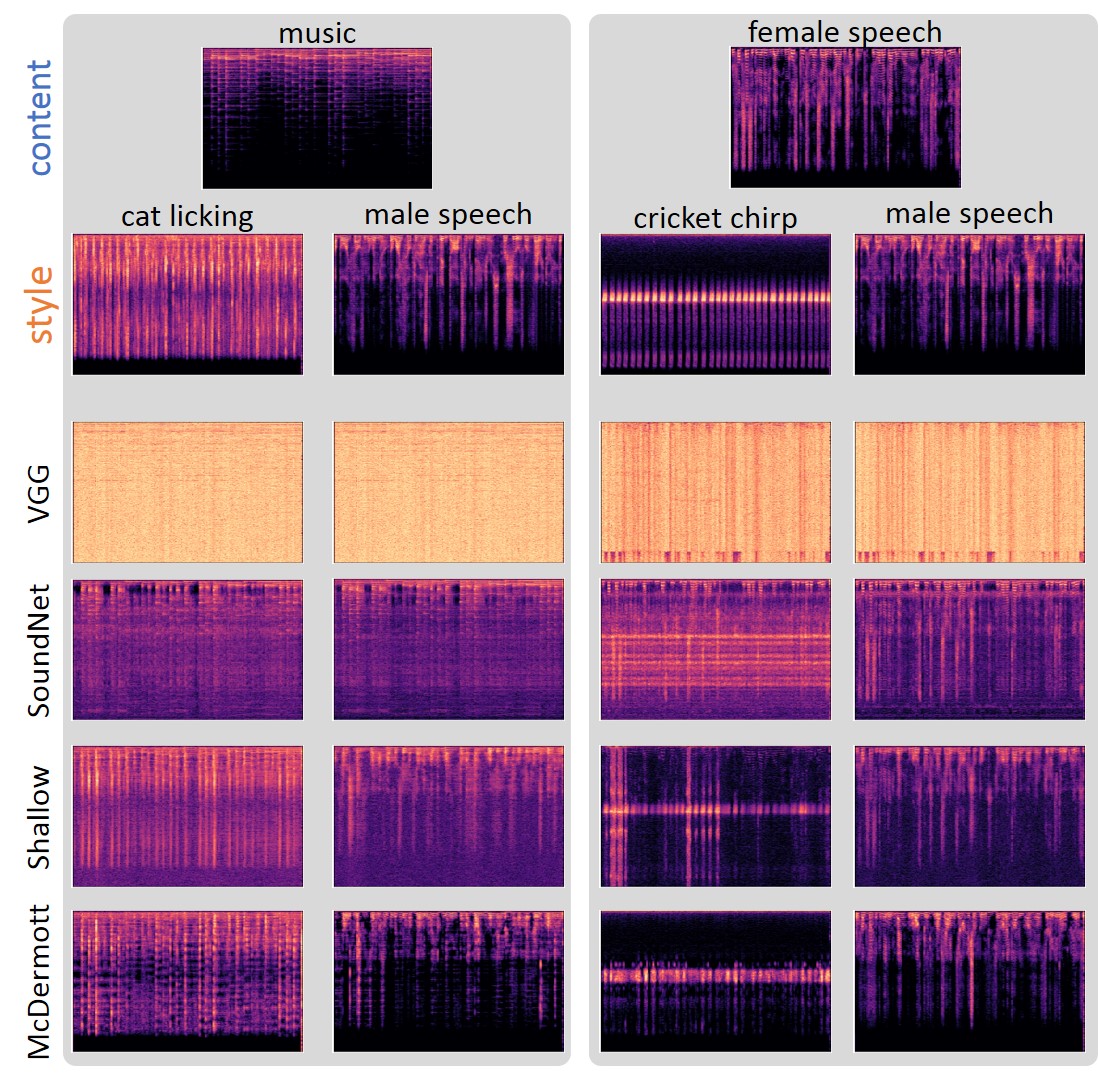} 
    \caption{\textbf{Result samples}: Spectrogram of the content sound (1st row), of the style sound (2nd row), and of the results obtained by different sound texture models (3rd to 6th rows).}
    \label{fig:result}%
\end{figure}

We evaluated the proposed framework using the four different sound texture models presented in previous section -- VGG-19 CNN, SoundNet CNN, Shallow CNN (short for wide, shallow and random) and McDermott auditory model --, with several types of content and style sounds. Our subjective tests have confirmed that initializing the optimization by the content sound itself yields much more meaningful results compared to the conventional workflow where a random noise signal is used instead in conjunction with a content loss. The results obtained by the latter approach are actually inline with the observations in \cite{ulyanov2016blog}. Fig. \ref{fig:result} shows the spectrograms of some content sounds (piano piece of Pachelbel's Canon in D  and speech of a woman speaking French), of the style sounds (male speech and simple textures like cat licking milk and crickets chirping), and of the results obtained by the four texture models. The corresponding audio files together with few more examples are available for informal listening on the supporting webpage.\footnote{\url{https://egrinstein.github.io/2017/10/25/ast.html}}

As can be seen and heard from the examples, the results obtained by VGG are extremely noisy and far from our expectations. This is comprehensible since VGG is designed and trained for image description. SoundNet offers some meaningful results, but the produced sounds contain a substantial noise and they seem not appealing in terms of style insertion. Surprisingly, the shallow net with random filters works far better than SoundNet and in many cases it succeeds in inserting local texture from the style sound into the global structure of the content sound. McDermott's model, which is designed to model human auditory system, produces compelling results. To better understand and interpret the results, one can see in the last row of Fig. \ref{fig:result} that in the case of music content, the attacks of piano notes are preserved in the transformed signal while the local texture of the style sound is introduced in synchronicity with these attacks. Using the ``cat licking milk'' as style and music as content, one can hear a ``cat lick'' at every onset of the piano notes. Finally, when using speech as style and piano music as content, one can feel as if the melody is hummed by the speaker even though the phonemes are lost. One can also hear and see that while the shallow net seems to place the style texture in the attacks more precisely, McDermott's model recreates better the local texture. Again in the example of the ``cat licking milk'' style, individual licks are placed with better precision for the shallow net, but we perceive them more as whips rather than licks. On the other hand, McDermott's model places licks less frequently in the resulting sound, but they are more recognizable by listening.

\section{CONCLUSION}
\label{sec: conclusion}

In this work, we have proposed a general framework for audio style transfer. For extracting the required audio texture statistics, we investigated the use of different neural network architectures as well as of a handcrafted model inspired by the human auditory system. Experiments showed that the latter and a shallow random net were the two able to produce promising results. Importantly, we found that initializing the iterative optimization by the content sound itself, and discarding the content loss, plays an important role in obtaining these results. This research topic is nascent and, at this stage, numerous aspects related to the design of relevant sound texture models should be further investigated in future work. Also the fundamental question of \emph{``what should be the style to transfer and what should be the content to preserve?''} is still largely open for future discussion.

\clearpage
\bibliographystyle{IEEEbib}
\balance
\bibliography{refs}

\end{document}